\documentstyle[aps,rotate,multicol,epsf]{revtex}
\def\beginwide{
        \end{multicols} \vspace*{-0.5cm} \noindent
        \rule{3.5in}{.1mm}\rule{.1mm}{5mm} \widetext \medskip }
\def\beginwidetop{
        \end{multicols} \vspace*{-0.5cm} \noindent
        \widetext \medskip }
\def\endwide{
        \hspace*{3.35in}~\rule[-5mm]{.1mm}{5mm}\rule{3.5in}{.1mm}
        \begin{multicols}{2} \vspace*{-1.0cm} \noindent }
\def\endwidebottom{
        \begin{multicols}{2} \vspace*{-1.0cm} \noindent }

\draft

\begin{document}
\title{Multiscaling and Structure Functions 
in Turbulence: An Alternative Approach
} 
\author{Mogens H. Jensen$^*$
}
\address{Niels Bohr Institute 
and Center for Chaos and Turbulence Studies, 
Blegdamsvej 17, DK-2100 Copenhagen {\O}, Denmark }
\date{January 15, 1999}
\maketitle
\begin{abstract}
We propose an alternative formulation of structure functions
for the velocity field in fully developed turbulence. Instead of averaging 
moments of the velocity differences as a function of the distance, 
we suggest to average moments of the distances as a function of
the velocity difference. This is like an ``inverted'' structure 
function, with a different statistics. On the basis of shell model calculations
we obtain a new multiscaling spectrum.
\end{abstract}
\begin{multicols}{1}
\smallskip
\smallskip
The understanding of intermittency effects in fully developed turbulence
and the associated multiscaling spectrum of exponents, is probably the most
fundamental open problem in turbulence research \cite{frisch}. The
traditional way of describing this is, as already suggested
by Kolmogorov \cite{Kol}, to consider the velocity difference between
two points of the turbulent state, raise this difference to the
moment $q$, and then study the variation with respect to the distance
between the two points. To improve the statistics, the moments 
are averaged in space (and maybe time)
and one obtains the well known structure functions where the corresponding
scaling exponents are called structure function exponents \cite{frisch}. 
During the
last decades it has become clear both from many experimental
\cite{Anselmet,Sreeni,vdWater}, numerical \cite{Meneguzzi} and 
theoretical considerations \cite{PV,Lvov},
that this set of exponents is very non-trivial, defining an infinity
of independent exponents leading to at ``curved'' variation of the
scaling exponent with the moment. Notable is also the recent
fundamental advances in obtaining the multiscaling spectrum analytically
for a passive scalar advection in a spatially correlated, but 
temporally non-correlated, velocity fields, the socalled Kraichnan
model \cite{Kupi,Falk} . 

We propose simply to ``invert'' the structure function equation,
and consider instead averaged moments of the distance between
two points, given a velocity difference between those points. This
leads to an alternative way of describing and analyzing a turbulent
velocity field (in particular when measured experimentally) 
and one obtains a new set of exponents that we have not
yet been able to relate to the traditionally estimated exponents, though
we suspect that there might be a relation. 
This inversion is inspired
by studies in passive scalar advection where one often, say for pair
particles, considers averages of the advection time versus the
distance, instead of averages of the distance versus 
time \cite{ABCCV,Massimo,Gat,BCCV}.
To a start let us introduce
the well known structure functions for the velocity 
field ${\bf u} ({\bf x}, t)$ of a fully developed turbulent state,
obtained either from the Navier-Stokes equations or from measurements
\begin{equation}
\label{vel}
<  \Delta  u_{\bf x} (\ell)^q > \sim  \ell^{\zeta_q}
\end{equation}
where the difference is defined as
\begin{equation}
\label{ua}
\Delta  u_{\bf x} (\ell) ~=~ {\bf u} ({\bf x + r}) - {\bf u} ({\bf x})~~,~~
 \ell = | {\bf r} |
\end{equation}
The average in Eq.(\ref{vel}) is over space (and maybe time) . We have assumed
full isotropy of the velocity field. The set of
exponents $\zeta_q$ forms a multiscaling spectrum
\cite{PV}.

Alternatively, we now consider the following quantities, which
is denoted the {\it distance structure functions}
\begin{equation}
\label{ds}
<  \ell (\Delta  u_{\bf x} )^q > \sim | \Delta u_{\bf x}|^{\delta_q}
\end{equation}
where the difference $\Delta u_{\bf x}$ is again defined as in
Eq. (\ref{ua}) and $\ell (\Delta u_{\bf x})$ is understood as the {\it minimal}
distance in $\bf r$, measured from $\bf x$, for which 
the velocity difference exceeds the value $\Delta u_{\bf x}$. In
other words, we fix a certain set of values of the velocity
difference $\Delta u_{\bf x}$. Starting out from the point $\bf x$,
we monitor the distances $\ell (\Delta u_{\bf x})$ where the
velocity differences are equal to the prescribed values. Performing
an average over space (and maybe time) the distance
structure functions Eq. (\ref{ds}) are obtained. 
By assuming self-similarity of the small scale velocity differences,
one expects a trivial set of exponents $\delta_q$
where the variation with the moment $q$ is determined by one exponent.
Say, in the standard Kolmogorov theory we know that the
velocity differences behave as $\Delta u \sim \ell^{1/3}$, forgetting
for a moment the averaging brackets. Inverting this equations, we
of course obtain $\ell \sim \Delta u^3$ and would expect a
trivial relation $\delta_q = 3 q$. In case of an
intermittent and singular velocity field without self-similarity of
the small scale velocity differences (see \cite{Frisch2}), 
this would be completely
different and the averaging brackets will be crucial, relating
to the statistics of the varying quantity that is averaged. We will
show, based on shell model calculations, that in turbulence there exists
a new spectrum $\delta_q$, that appears not to be trivially
related to the spectrum $\zeta_q$ \cite{Ver,Biferale}. 
Let us for a moment reflect
on the case $q=1$. 
Using the standard value $\zeta_1 \sim 0.38-0.40$, the 
simple inversion gives $\delta_1 \sim 2.5$.
Our calculations indicate that this value is not obtained in
a turbulent model field.
Instead we find a value $\delta_1 \sim 2.0-2.1$. Another way to do
the comparison is to aim at velocity exponent 1 and find the 
corresponding moment $\hat q$, i.e. $< \ell^{\hat q} > \sim
\Delta u^1$. We obtain $\hat q \sim 0.45$, again different
from 0.40.
These differences are of course attributed to
the very different statistics, i.e. whether the
velocity differences or the corresponding distances are averaged.
Also, we obtain strong intermittency corrections
in the sense that the value of the 8'th moment 
is $\zeta_{8} \sim 12.9$, i.e. much smaller than $8 \zeta_1 \sim 16.3$.

In order to apply this scheme in a direct calculation we employ
the Gledzer-Ohkitani-Yamada, GOY, shell 
model \cite {Gledzer,OY} which has
be intensively studied over the last 
years \cite{JPV,pisarenko,benzi,kada1,bif,kada2,Lohse}. This model is
a rough approximation to the Navier-Stokes equations and is
formulated on a discrete set of $k$-values, $k_n=r^n$.
We use the standard value $r =2$.
In term of a complex Fourier mode, $u_n$, of the velocity field 
the model reads
\begin{eqnarray}
\label{un}
(\frac{d}{ dt}+\nu k_n^2 ) \ u_n \ & = &
 i \,k_n (a_n \,   u^*_{n+1} u^*_{n+2} \, + \, \frac{b_n}{2} 
u^*_{n-1} u^*_{n+1} \, + \, \nonumber \\
& & \frac{c_n }{4} \,   u^*_{n-1} u^*_{n-2})  \ + \ f \delta_{n,4},
\end{eqnarray}
with boundary conditions $b_1=b_N=c_1=c_2=a_{N-1}=a_N=0$.
$f$  is an external, constant forcing, here on the forth mode.
\begin{figure}[htb]
\narrowtext
\vbox{
{
        \centerline{\epsfxsize=5.0cm
        \rotate[r]{\epsfbox{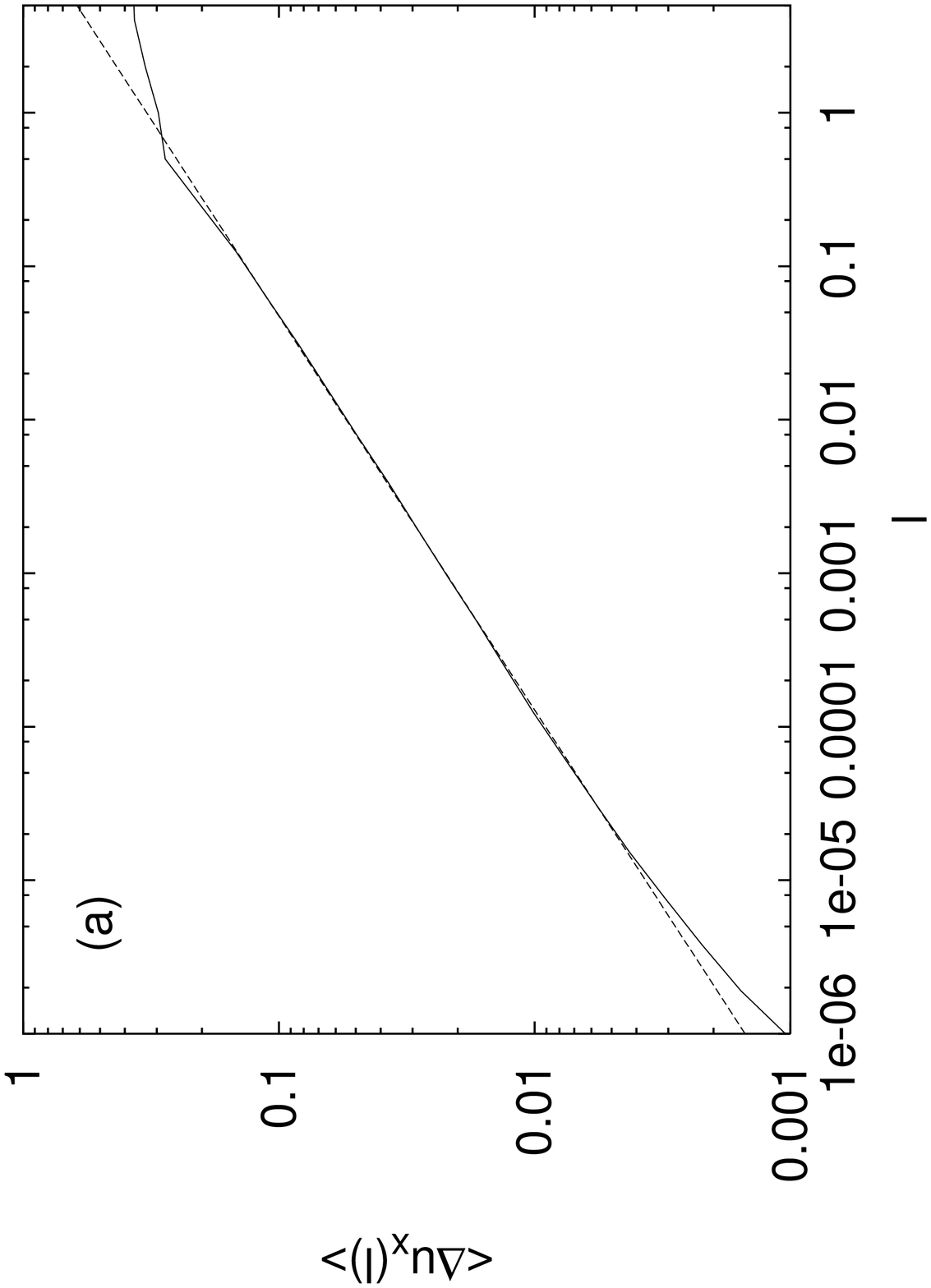}}}
}
{
        \centerline{\epsfxsize=5.0cm
        \rotate[r]{\epsfbox{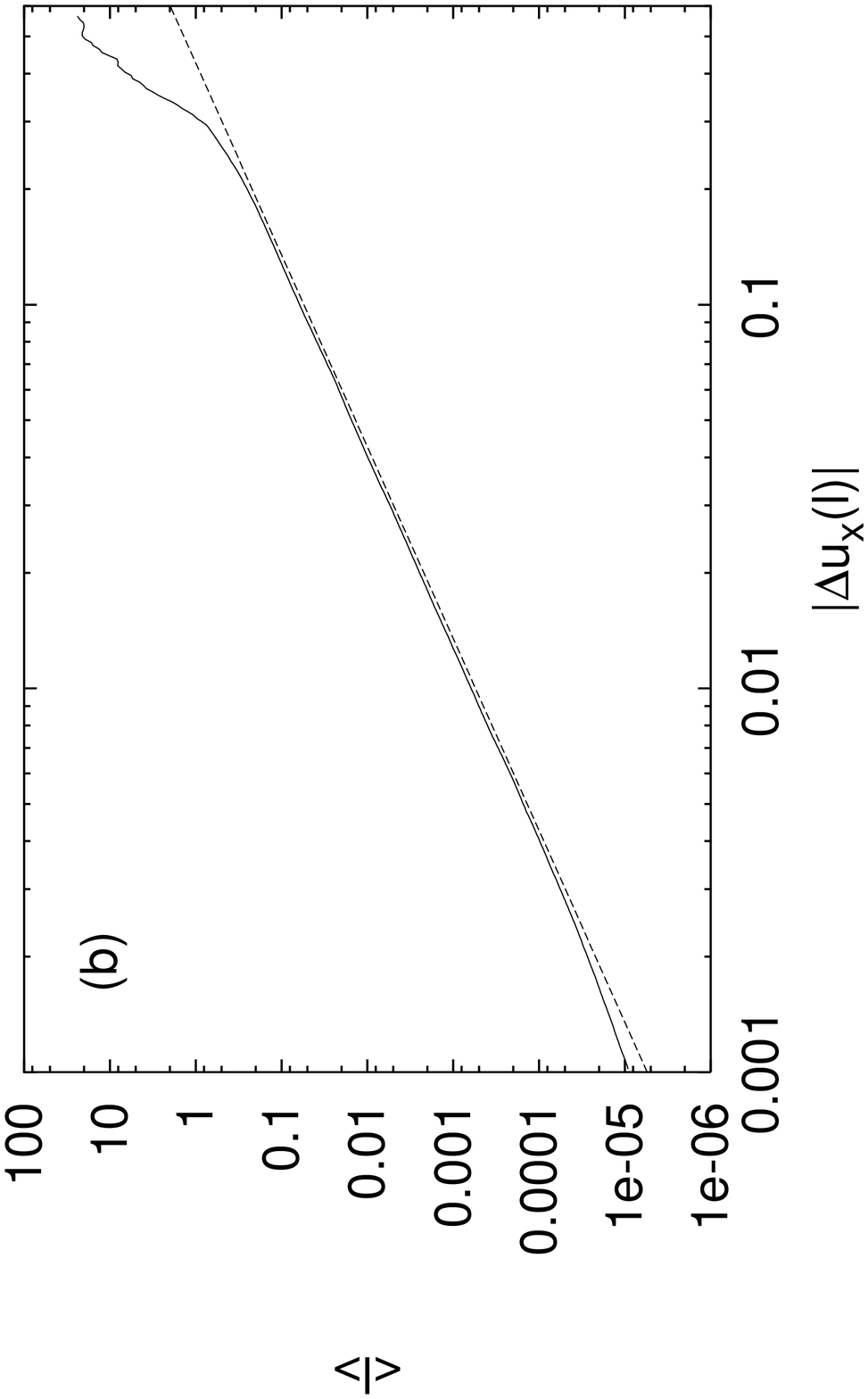}}}
}
{
        \centerline{\epsfxsize=5.0cm
        \rotate[r]{\epsfbox{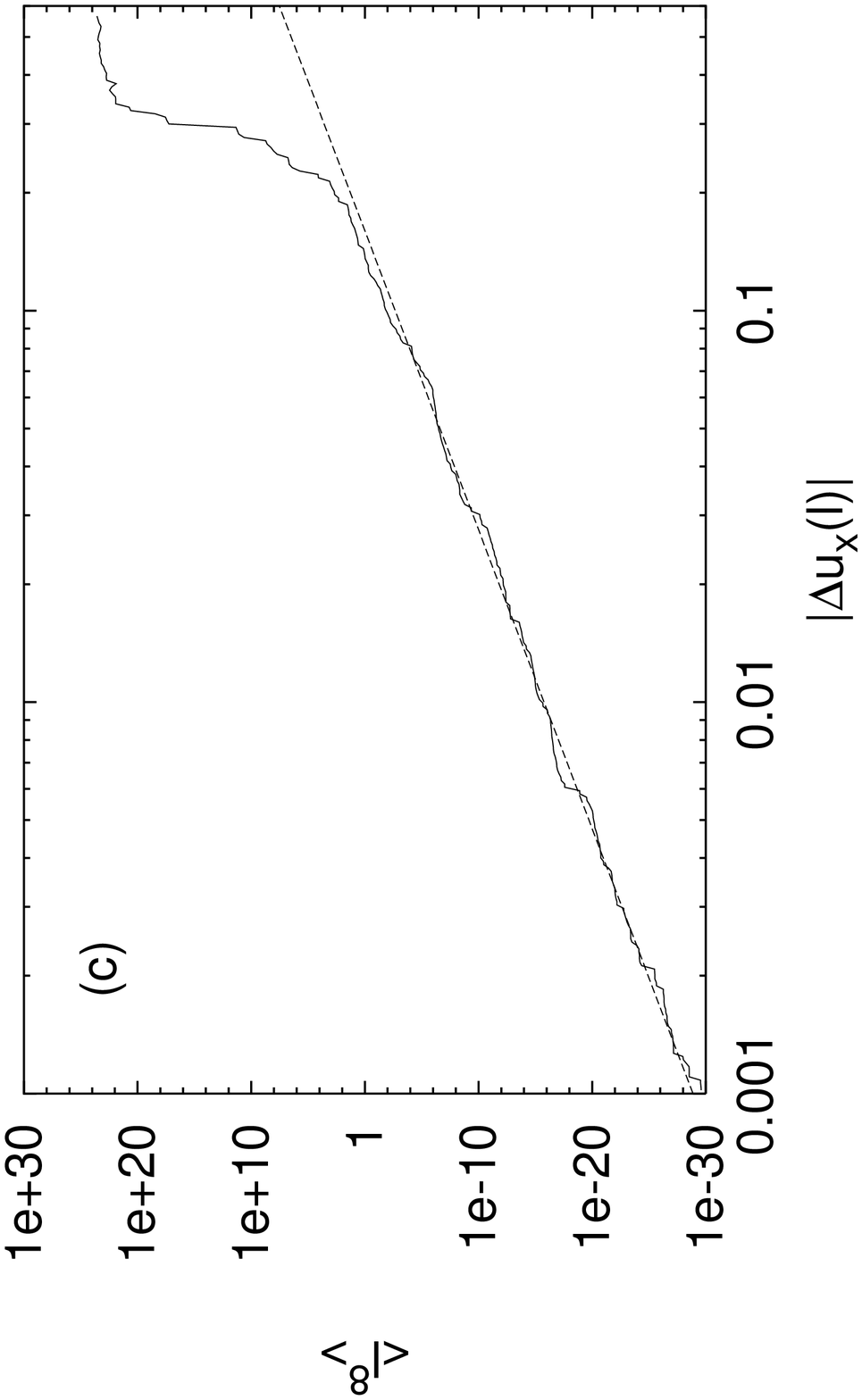}}}
}
}
\caption{
a): The velocity structure function of order one. The
line has a slope of 0.39.
b): The distance structure function of order one. The line has
a slope of 2.02. Note the inner cut-off related to the dissipative
cut-off in a),
and the outer cut-off given by velocity of the forcing scale. c):
The distance structure function of order 8. The exponent is 
$\delta_8 \sim 13.1$. The ``raggedness'' is due to discretization of the varying
length scale.
}
\end{figure}

The  coefficients of the non-linear terms  must follow the relation
$a_n+b_{n+1}+c_{n+2}=0$ in order  to satisfy the conservation  of  energy,
$E = \sum_n |u_n|^2$, when $f=\nu = 0$. 
The constraints still leave a free parameter $\epsilon$ so that 
 one can set
$ a_n=1,\ b_{n+1}=-\epsilon,\ c_{n+2}=-(1-\epsilon)$ \cite{bif}. As observed
by Kadanoff, one obtains the canonical value $\epsilon= 1/2$, if 
helicity conservation is also demanded \cite{kada1}. 
The set (\ref{un}) of $N$ coupled
ordinary differential equations can be numerically integrated by
standard techniques.
We have used standard parameters in this paper $N = 27, \nu = 10^{-9},
k_0 = 0.05, f = 5 \cdot 10^{-3}$.
\begin{figure}[htb]
\narrowtext
\vbox{
{
        \centerline{\epsfxsize=5.0cm
        \rotate[r]{\epsfbox{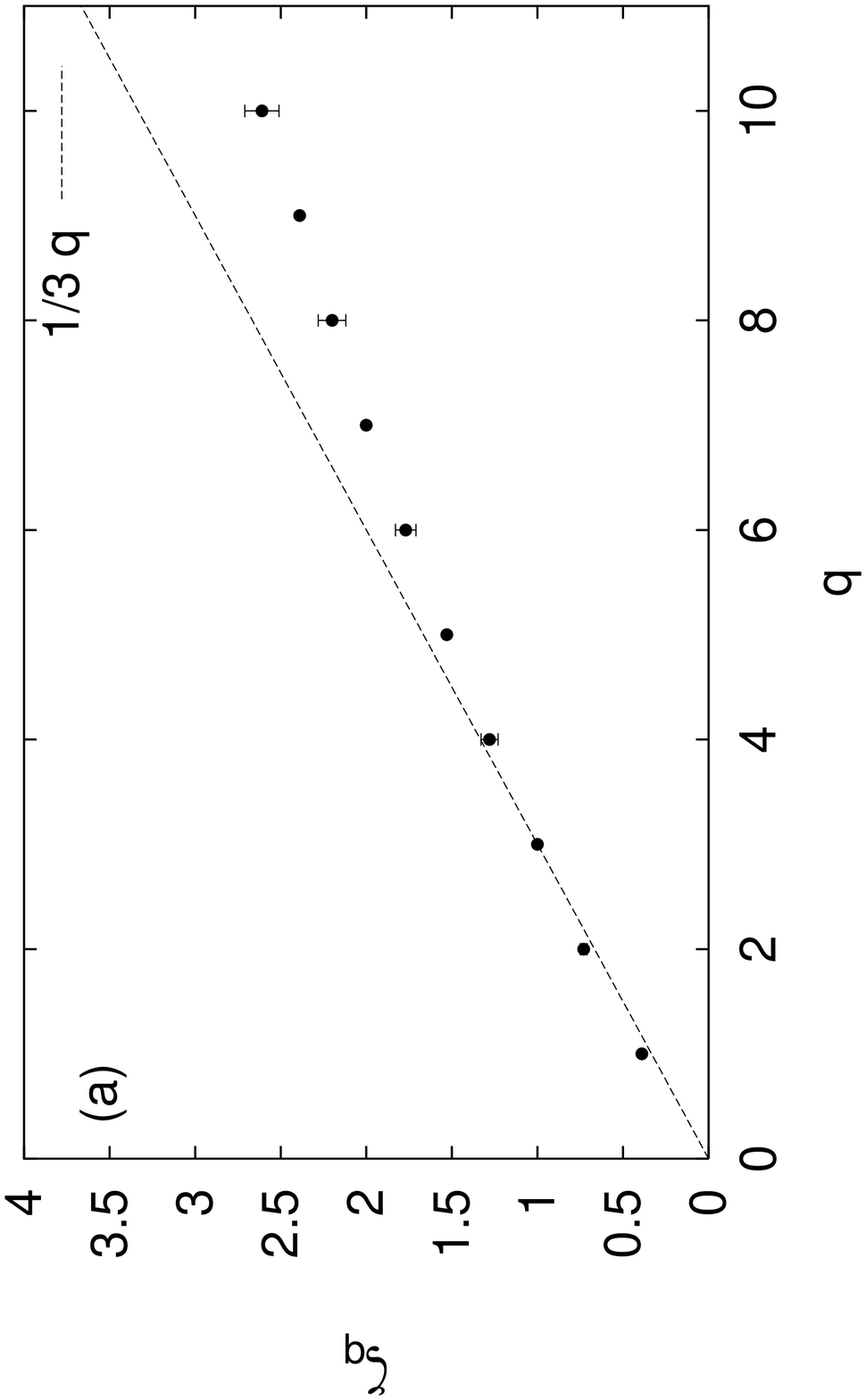}}}
}
{
        \centerline{\epsfxsize=5.0cm
        \rotate[r]{\epsfbox{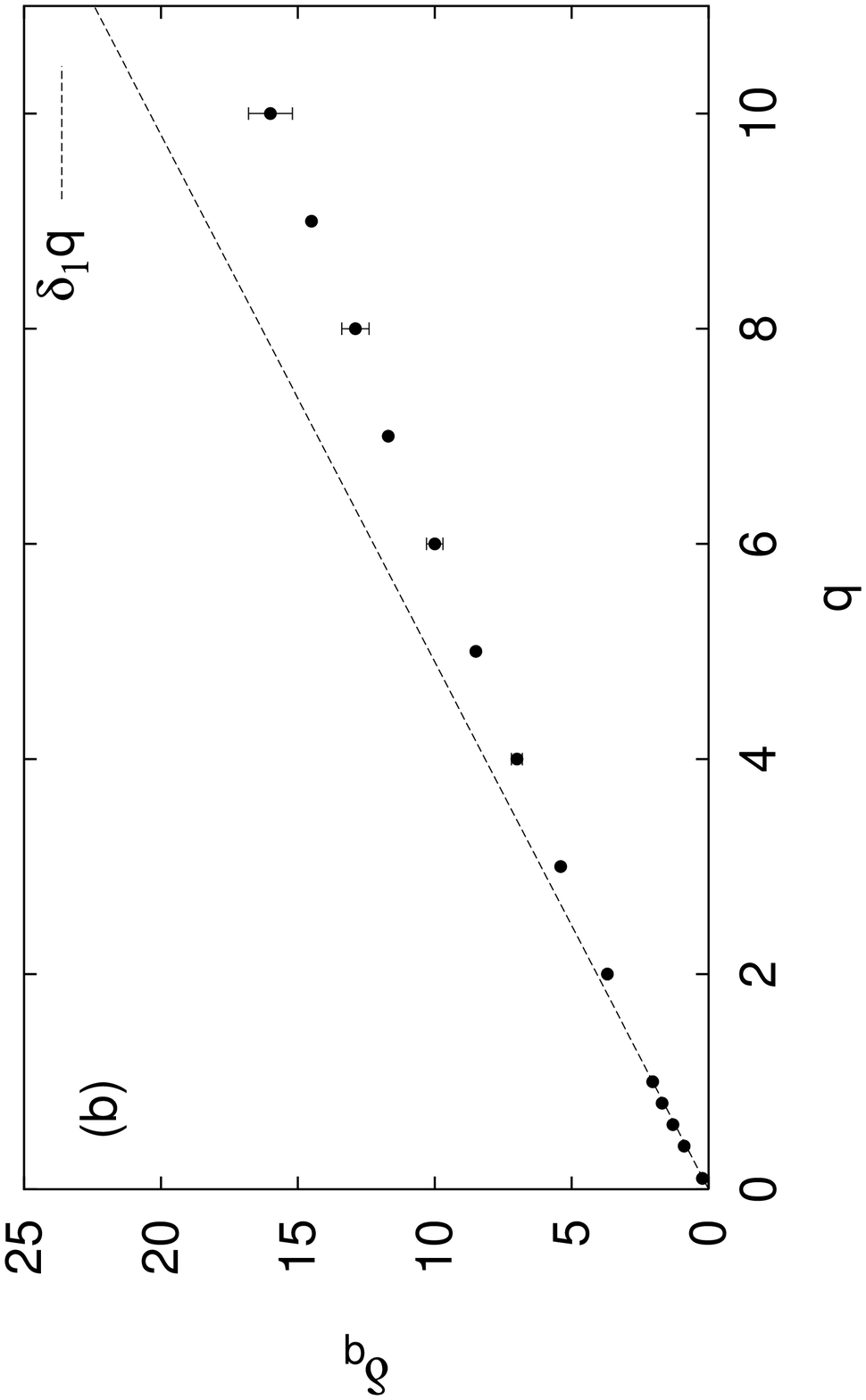}}}
}
}
\caption{
a): The exponents $\zeta_q$ for the velocity structure functions,
with selected error bars. The line corresponds to Kolmogorov theory.
b): The exponents $\delta_q$ for the distance
structure functions. The line is adjusted to pass through the value of the first
order exponent $(1,\delta_1)$.
        }
\end{figure}

The GOY model is defined in $k$-space but our formalism
is written in direct space and we therefore apply a sort of
inverse Fourier transform \cite{note}. Here we employ an idea proposed 
by Vulpiani \cite{AP,mogens} and write the three-dimensional
velocity field in the following way
\begin{equation}
\label{field}
 {\bf u} ({\bf x},t)=\sum_{n=1}^{N} 
 {\bf c}_n [u_n(t) e^{i {\bf k}_n\cdot {\bf x}}
+ c.\, c.]
\end{equation}
The wavevectors are defined by
\begin{equation}
{\bf k}_n ~=~ k_n {\bf e}_n
\end{equation}
where ${\bf e}_n$ is a unit vector in a random direction, for each
shell $n$. 
Also ${\bf c}_n$ are unit vectors in random
directions. One can easily ensure that the velocity field
is incompressible, $div ~ {\bf u} =0$, by the following
constraint \cite{AP}
\begin{equation}
{\bf c}_n \cdot {\bf e}_n =0 ~~~~ \forall n ~~~~.
\end{equation}
Note, that this condition could be relaxed to
$\sum_{n=1}^N {\bf c}_n \cdot {\bf e}_n =0$. In our numerical
computations we consider the vectors
${\bf c}_n$ and ${\bf e}_n$ quenched in time but nevertheless
average over many different realizations of these; i.e. 
one, or several, specific measurements of the distance structure
functions are performed with one realization of the vectors. 
After that a new
realization of ${\bf e}_n, {\bf c}_n$ is applied in order to perform
a good statistical average.
\begin{figure}[htb]
\narrowtext
\vbox{
{
        \centerline{\epsfxsize=5.0cm
        \rotate[r]{\epsfbox{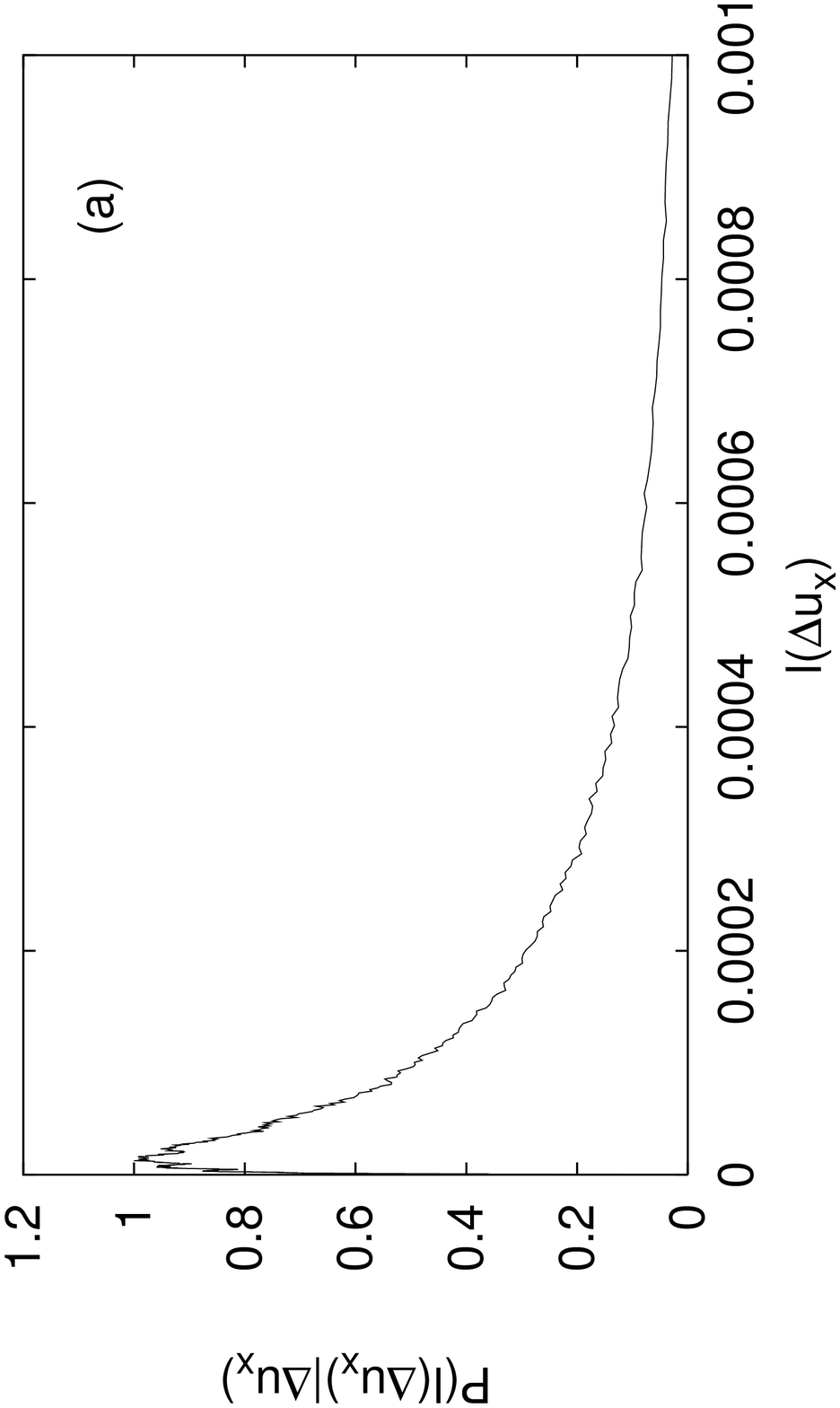}}}
}
{
        \centerline{\epsfxsize=5.0cm
        \rotate[r]{\epsfbox{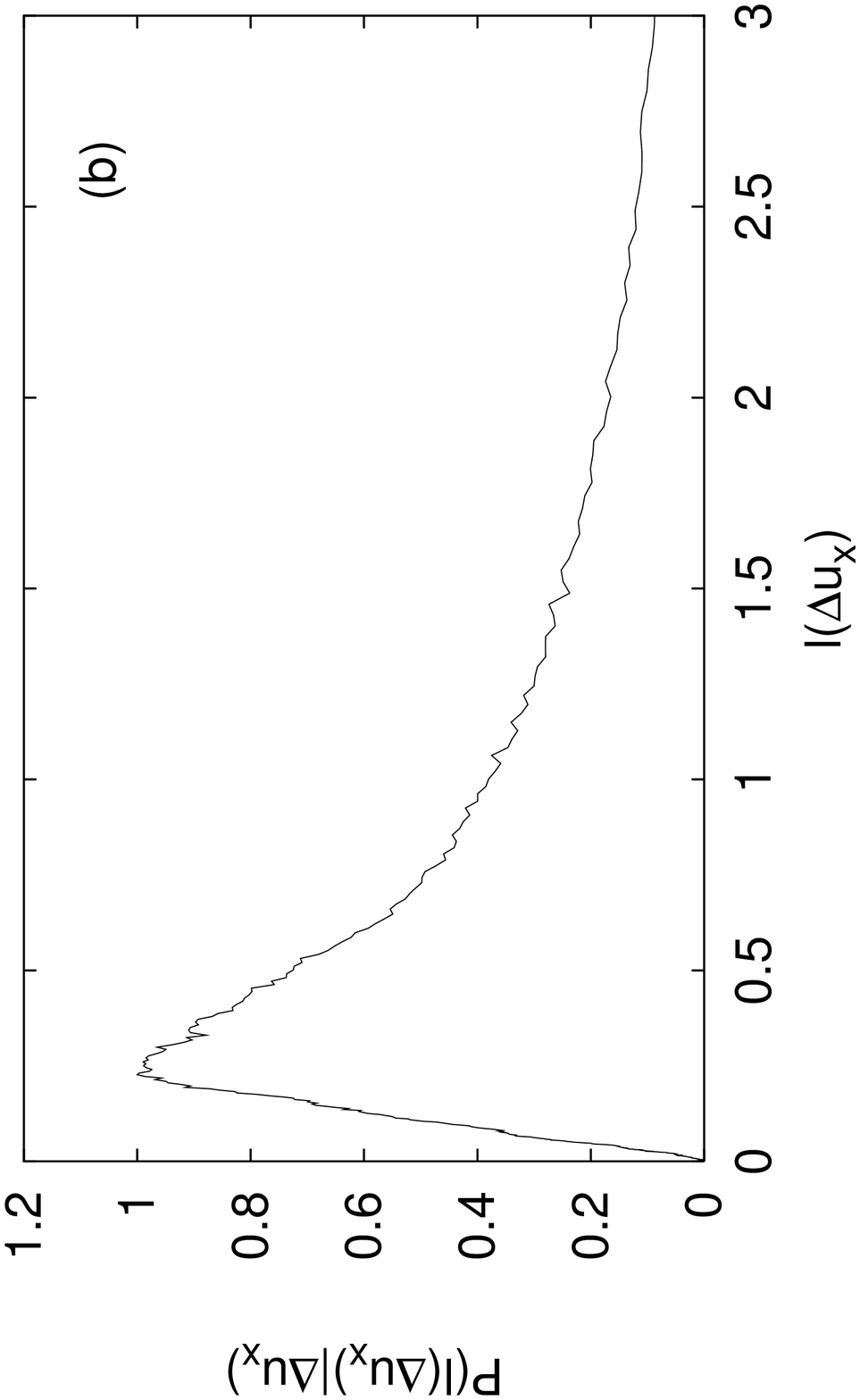}}}
}
}
\caption{
Probability distribution functions
$P(\ell (\Delta u_{\bf x}) | \Delta u_{\bf x} )$ for (a) the
velocity $\Delta u_{\bf x} = 0.0027$, which is close to the dissipative
length scale (see Fig.1), and for (b) the
velocity $\Delta u_{\bf x} = 0.26$, close to the velocity of
the outer cut-off.
        }
\end{figure}

Equipped with a real space time dependent velocity field we start out
with a test of this field by computing the standard velocity structure
functions, given by Eq. (\ref{vel}) \cite{note1}.
Indeed, the field exhibits nice
scaling invariance as shown in Fig.1a, where the first order velocity
structure function
is presented. We have extracted all the exponents with moment
up to $q$=10 and the corresponding results are shown in Fig.2a
(and for completeness also in Table 1). 
These results agree with the exponents obtained 
by numerical computations of the GOY-model
in $k$-space \cite{JPV}, i.e. without performing
the transformation to real space. In the averaging, we have assumed
isotropy and for practical convenience, the distance is varied
only along the three coordinate axes.  Having checked this we
proceed to extract the distance structure functions, Eq. (\ref{ds}).
Practically, both the distance and velocity differences are discretized
as $\ell = \lambda_d^i$ and 
$\Delta u = \lambda_u^j$. In the present calculations 
the value $\lambda_d = \lambda_u = 1.02$ is chosen.  As the starting
point we set $\bf x = 0$ and vary again along the 
coordinate axes. For a fixed value of $\Delta u_{\bf 0}$, 
$\ell$ is increased until {\it for the first time} the velocity
difference exceeds this fixed value: this defines 
$\ell (\Delta_{\bf 0} u)$. Then $\Delta u_{\bf 0}$ is increased
by one more step and so on. Fig.1b presents the scaling of the
first order distance structure function and the corresponding 
exponent $\delta_1$ is estimated to a rather good
precision, $\delta_1 = 2.02 \pm 0.05$, with a scaling regime of 2
decades on the $\Delta_{\bf 0} u$ axes and 4-5 decades on the
$\ell$ axes. Note, the cut-off at low values of $\Delta_{\bf 0} u$. This
cut-off is related, both for values of velocity and distance,
to the dissipative cut-off of the
standard structure function, see Fig. 1a.
The cut-off at large values of $\Delta_{\bf 0} u$ is 
related to the velocity at the forcing scale. In all the presented
calculations we have averaged over 24630 situations.
Fig. 1c presents the distance
structure function of order $q=8$, resulting in an exponent
$\delta_8 = 12.9 \pm 0.5$. The graph is ``rough'' due to the binning
of $\Delta_{\bf 0} u$ and due to the high moment. Fig. 2b shows the 
multiscaling spectrum of $\delta_q$. We have included a straight line
through the point $(1,\delta_1)$ in order to show the curved
nature of the spectrum. For completeness, 
Table 1 also displays the measured scaling
exponents $\delta_q$. 

It is well known, that one can improved the
scaling significantly using the technique of extended self 
similarity (ESS) \cite{Cili} where one moment of a given variable is varied
against another moment. In the present case 
this means a graph of one distance structure function
$< \ell (\Delta_{\bf 0})^q >$ versus another $< \ell (\Delta_{\bf 0})^{q'} >$ 
for two different moments $q,q'$, and this
results in ESS plots which spans over three times as long a
regime as compared to traditional ESS plots where the
quantities are $<  \Delta u_{\bf x} (\ell)^q >$ are applied
(the large regime is of course due to the Kolmogorov $\frac{1}{3}$
exponent relation). Applying ESS we have obtained
the exponents $\delta_q$ in an independent way and the results 
agree well with the values listed in Table 1. 
This property
of a much larger scaling regime of the ESS plots could be
one of the advantages of the presented formalism. Details
will be given in a forthcoming publication.

To obtain a better understanding of the obtained results we
need to consider the statistics of
$ \ell (\Delta u_{\bf x})^q $ in the following way
\begin{equation}
< \ell (\Delta u_{\bf x})^q > \simeq \int~ \ell (\Delta u_{\bf x})^q
P(\ell (\Delta u_{\bf x}) | \Delta u_{\bf x} )~ d \ell
\end{equation}
where we have introduced the {\it conditional probability 
distribution function} $P(\ell (\Delta u_{\bf x}) | \Delta u_{\bf x} )$. 
This measures the probability of a distance $\ell$ given the 
velocity difference. We show this PDF for two different values
of the velocity difference in Fig. 3 on linear scales.
In both cases, the distributions are clearly non-Gaussian with
long exponential (or in fact stretched exponential) tails,
as expected in intermittent systems. The surprising difference
to the standard PDF's for velocity differences is, that it does
not tend towards a Gaussian for large scales. We would have expected that. 
We have not been able to relate
this PDF, $P(\ell (\Delta u_{\bf x}) | \Delta u_{\bf x} )$, to the
``usual'' PDF, $P( \Delta u_{\bf x} | \ell )$; these two PDF's
measure simply very different things.

In conclusion, we have introduced the distance structure
functions defined for a velocity field in fully developed
turbulence. The corresponding multiscaling spectrum appears
not to be related to the well known spectrum for velocity
structure functions. The distance structure function could
be very relevant for experimental velocity data measured in
one point \cite{Biferale}. Here one typically applies the Taylor hypothesis
in order to relate a temporal segment to a spatial segment.
For this type of time series, the distance structure functions
should be easily extracted.

I am grateful to J. Sparre Andersen, L. Biferale, P. 
Muratore-Ginanneschi, M. Vergassola and A. Vulpiani for comments
and discussions.

\begin{table}[htb]
\caption{Values of the scaling exponents for velocity structure
functions $\zeta_q$ and distance structure functions $\delta_q$
with selected error bars. 
        }
\vspace*{2mm}
\begin{tabular}{||c|c|c|c|c|c|c|c|c|} 
q  & 1  &  2  & 3 & 4 & 5 & 6 &7& 8 \\
\hline
$~\zeta_q~$ & 0.39 $$& 0.73(2) & 1.0 $$& 1.28(5) & 1.53 $$& 1.77(6) & 2.0 $$& 2.20(8)\\
$~\delta_q~$ & 2.04 $$& 3.70(5) & 5.4 $$& 7.0(2) & 8.53 $$& 10.0(4) & 11.7 $$& 12.9(6)\\
\end{tabular}
\vspace*{0.5cm}
\label{table-exp}
\end{table}
\end{multicols}


\begin{thebibliography}{10}
\bibitem{note}
We have also tested the formalism in $k$-space but
obtain less good statistics compared to the real space calculation.
Nevertheless, by introducing a large number of shells which are less
separated, say $N \sim 1000$ and $k_n = 1.05^n$, the k-space
computation can be pursued with good results.
\bibitem{note1}
The formalism has been applied to real experimental 
velocity fields using exit times in \cite{Biferale}.
\end{thebibliography}
\end{document}